\documentclass[aps,twocolumn,floatfix]{revtex4}

\usepackage{graphicx}

\begin{document}

\title{Materials Science and Protein Crystallography\\
	Using the MX Beamline Control Toolkit}
\author{W.\ M.\ Lavender}
\affiliation{Biological, Chemical, and Physical Sciences Department,
	Illinois Institute of Technology, Chicago, Illinois 60616}

\date{\today}

\begin{abstract}
MX is a portable beamline control system that has been described at previous
NOBUGS meetings. 
This paper will briefly review MX and then
discuss important changes and improvements made
since the last meeting in 2000.

For materials science, work has focused on extending 
the support for multichannel analyzers and for fast data
acquisition using quick scans.
MX MCA support has focused on the development of interfaces
to the X-Ray Instrumentation Associates DXP-2X and X10P (Saturn) MCAs.
The MX DXP-2X support has been used by MR-CAT at the Advanced Photon Source
to readout a 13-element Ge detector at input count rates of up to $1.5*10^6$
counts per second per detector channel.
The other major addition is support for quick scans using multichannel
scalers.
Quick scanning is now routinely used for XAFS and diffraction measurements
at MR-CAT and will soon be implemented on some MX crystallography beamlines
as well.
We have also begun work to allow XIA MCAs to be read out during quick scans.

For protein crystallography, we have primarily focused on implementing MX for
new beamlines, namely, SER-CAT at the APS and GCPCC at CAMD, with others
pending at the APS.  Progress has also been made on the integration
of MX with vendor CCD and robotics software.
\end{abstract}

\maketitle

\section{Introduction}
The MX beamline control toolkit is a general package for control of
synchrotron radiation beamlines and laboratory X-ray generator systems
that is being jointly developed by the 
MR-CAT\cite{MRCATwebsite}\cite{MRCAT},
IMCA-CAT\cite{IMCACATwebsite}\cite{IMCACAT1}\cite{IMCACAT2},
and SER-CAT\cite{SERCATwebsite}
sectors at the Advanced Photon Source.
It has been described at previous NOBUGS meetings
\cite{LavenderNOBUGS1997}
\cite{LavenderNOBUGS2000}
and elsewhere\cite{LavenderSRI99}.
This paper will focus on changes and additions to MX made since the last
NOBUGS meeting in 2000.

\section{Review of MX}

MX is a portable beamline control toolkit available from the web site
\url{http://www.imca.aps.anl.gov/mx/}.  Current users of MX are listed
in Table \ref{tab:Beamlines}.  The primary goal of MX is to make it 
possible to write beamline control system software that can be reused
in almost any environment.  This goal was conceived based on the author's
frustration with several previous control systems he had worked with that
paid little or no attention to transportability of the code.
\begin{table}
\begin{tabular}{|l|l|l|}
\hline
Group       &Experiment Type\ &Beamline Equipment\\
\hline
MR-CAT      &Material Science &EPICS motor, scaler, MCS,\\
(APS)       &                 &McLennan, Compumotor, \\
            &                 &Newport, XIA MCAs, etc.\\
\hline
IMCA-CAT    &Crystallography &EPICS motor, scaler, \\
(APS)       &                &McLennan, Compumotor, \\
            &                &Newport, etc.\\
\hline
SER-CAT     &Crystallography &Delta Tau PMAC, \\
(APS)       &                &Struck VME MCS\\
\hline
DND-CAT     &Crystallography &SCIPE motor, counter, \\
(APS)       &                &timer support, XIA MCA\\
\hline
GCPCC       &Crystallography &Compumotor 6K, \\
(CAMD)      &                &Ortec counter/timer, \\
            &                &XIA MCA\\
\hline
\end{tabular}
\caption{\label{tab:Beamlines}A list of beamlines currently using MX.}
\end{table}

MX attempts to achieve this goal in several ways:
\begin{itemize}
\item MX is designed to be easily ported to new operating systems.
	It already runs on Linux, Microsoft Win32, MacOS X, Solaris,
	Irix, HP/UX, and Cygwin with legacy support for SunOS 4, and MSDOS.
\item MX is written in ANSI C to maximize its ability to be invoked from
	other packages.  Most applications and scripting languages that
	support an external calling interface can easily invoke code
	written in C.
\item MX localizes dependencies on particular network protocols to relatively
	small areas of the code.  Thus, while MX comes with its own
	network protocol, it is designed to make it easy to use other
	network protocols instead.  For example, it is easy
	to use MX on a beamline that uses only 
	EPICS\cite{EPICS} network protocols.
\item The core of MX is designed as middleware that can be inserted into
	other systems as necessary.  This is done by providing a high
	level application programming interface (API) to applications
	or servers which tries to avoid making assumptions about the
	nature of those applications and servers.  At the low end,
	a device driver API is provided that tries to avoid making
	assumptions about the nature of the hardware underneath.
\end{itemize}

MX has been under development since 1995 and has developed a relatively
large number of device drivers.  Currently there are 34 drivers for
various motor controllers, devices that can be treated as motors, and
for access to motors controlled by other control systems such as EPICS.
There are also 17 pseudomotor drivers for items like X-ray energy,
monochromator control and goniostat table control.  In addition, there
are a variety of drivers for other devices such as scalers, timers,
MCAs, MCSs, and so forth with over 250 drivers in all.

\begin{figure}
\includegraphics{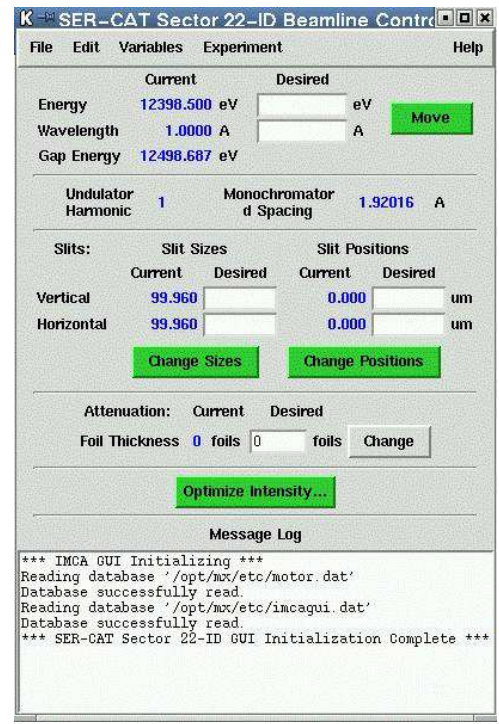}
\caption{\label{fig:Imcagui} The Imcagui protein crystallography program.}
\end{figure}

\begin{figure}
\includegraphics{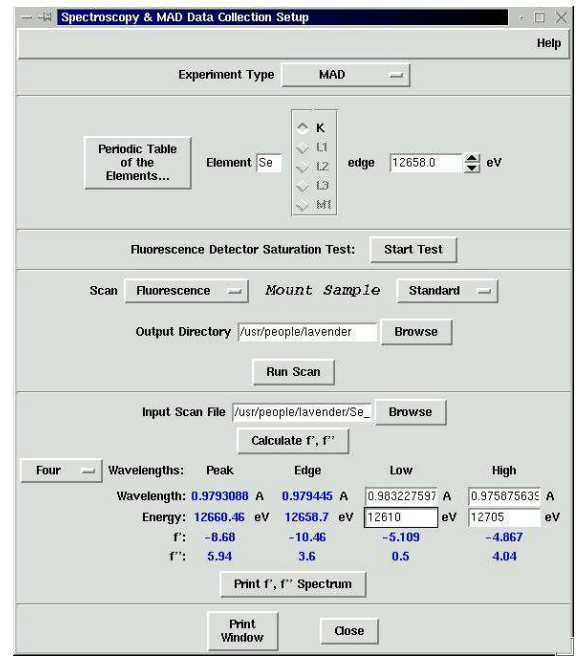}
\caption{\label{fig:ImcaguiMAD} Imcagui MAD experiment control window.}
\end{figure}

While MX can  be viewed as a set of components, it can also be used as
a complete beamline control system.  It provides an MX server that can
be used to manage beamline hardware and a set of graphical user interface
tools.  At present, MX supports developing GUI programs via its Python
and Tcl/Tk scripting interfaces.  The most important of these include:
\begin{itemize}
\item \textit{Imcagui} - A primary user interface for crystallography
	seen in Figure \ref{fig:Imcagui}.
	It provides support for monochromator, slit, and filter control.
	It also provides a way of executing fluorescence and absorption
	scans for multiwavelength anomalous diffraction (MAD) experiments
	as seen in Figure \ref{fig:ImcaguiMAD}
	which are then processed by Chooch to compute $f^{\prime}$
	and $f^{\prime\prime}$ values.  It was originally developed for
	IMCA-CAT but is now used at other beamlines as well.
\item \textit{Optimize} and \textit{optimize\_auto} -
	Automatic beamline intensity
	optimization programs that give beamline users a simplified
	interface for maximizing X-ray intensity on their own initiative.
\item \textit{Mxgui} - A more staff-oriented GUI for performing motor moves
	and executing arbitrary scans.
\end{itemize}
There are also a number of command line programs and utilities for MX.
The most important of these is \textit{motor} which is routinely used
at MR-CAT for materials science experiments.

Most of our existing GUI programs are written in Tcl/Tk.  However,
we now plan to write most new GUIs in Python instead.
Our experience has been that the syntax of Python seems more natural
to scientists than that of Tcl.  There also appears to be a groundswell
of support for Python at synchrotron radiation facilities as well,
given Python's extensive support for scientific and numerical programming.
With that in mind, we have now created an MX interface to Python
called MP and have begun writing applications using it.

\section{XIA Multichannel Analyzers}
A major task has been the development of software to interface with
the new high performance multichannel analyzers (MCAs) from
X-Ray Instrumentation Associates in Newark, California.
Our efforts have focused on the DXP-2X and the Saturn, formerly called
the DXP-X10P.

The DXP-2X is a CAMAC based system that contains four MCA channels
per CAMAC module.  The existing MX interface uses the Xerxes library
provided by XIA to control the DXP-2X.  The Xerxes library provides
direct access to many of the low level features of the MCA.  Since
the DXP-2X is a very powerful and complicated system, it took
a substantial amount of effort to correctly program the MX software
interface.  XIA now provides a new high-level software library called
Handel, but this was not available at the time of the original
development of our software.

There were a number of problems encountered in getting the DXP-2X
to run reliably, but these were partially due to the fact that the
DXP-2X firmware was still under development at the time.
Most of the problems we encountered have been solved now
and the DXP-2X owned by MR-CAT has been reliably used with a
13-element Ge detector from Canberra for several experimental runs now.
It has proved to be capable of handling much higher input count rates
than the other competing systems that were available at the time
of purchase.  The DXP-2X was recently used at MR-CAT to acquire data from a
sample of Pu(VI) adsorbed on MgO
at input count rates of up to $1.5*10^6$ counts per second per
MCA channel.

MR-CAT currently controls the DXP-2X via an MX server running on
a Windows 98 computers.  Since the XIA staff currently do all of their
development and testing on Windows platforms, it was regarded as desirable
to use the vendor library as provided to maximize XIA's ability to
diagnose problems.
Most of the rest of the beamline runs under Linux, so we anticipate
moving control of the XIA MCAs over to Linux at some point.
If it is moved to the Linux user interface computer, this will also
improve performance by eliminating a network transfer that must otherwise
be done.

Another XIA multichannel analyzer that we are interested in is the Saturn,
which was formerly known as the DXP-X10P.  This is a single channel MCA
that is controlled via a PC parallel port.  The Xerxes library used with
the DXP-2X also supports the Saturn.  Thus, after a few initial difficulties,
we were able to get the MX driver to work with the Saturn as well.
An optional feature of the Saturn that is not available for the DXP-2X
is 16 TTL output connectors that are associated with the
16 regions of interest (ROIs) of the Saturn.  In this configuration,
when a detector pulse falls within the boundary of a region of interest,
a TTL pulse is synthesized on the corresponding output channel.
Thus, in this mode one can think of the Saturn as a set of
16 programmable single channel analyzers whose output can be fed
into conventional pulse-counting scaler/timer systems.
This is the mode that DND-CAT currently uses.

\section{Quick Scans}

Another major effort over the last two years has been the development
of software to support quick scans, also known as slew scans or fast scans.
This was under initial development at the time of the last NOBUGS
meeting and has been routinely used now for two years at MR-CAT.

Quick scans are designed as a direct replacement for many types of step scans.
The problem with step scans is that there is a large delay associated with
slowing down the motor to a stop to take the data point and then 
accelerating back to speed.  Quick scans avoid this issue by acquiring
data continuously while the motors are in motion.
Quick scans are available for most motor drivers and for many of the
pseudomotor drivers as well.  For example, it is possible to quick scan
monochromator energy pseudomotors.

Of course, quick scans need a place to store the data as it is acquired.
This role is normally played by a multichannel scaler (MCS).  A multichannel
scaler counts pulses from a set of inputs and then, on receipt of a 
channel advance pulse, writes the counter values to a buffer and then
starts acquiring again.  This is then repeated for each data point
and at the end of the scan, the buffer containing all of the measurements
is read out.

For the data to be useful, we must also record the motor positions for
each point in the scan.  This is most easily done by recording some
signal that depends on the motor position.  The most obvious candidate
is a quadrature encoder signal from the motor controller that is slaved to
the position of the motor.  At MR-CAT, this is handled by taking
the encoder signals into a small electronics module that converts the
signals into two pulse trains.  One pulse train corresponds to motion
in the positive direction, while the second pulse train corresponds
to motion in the negative direction.
If the motor controller starts the move a few measurements after the
MCS has started acquiring data, then the absolute position of the motor
can be determined by taking the difference between the pulse count
in the positive motion channel and the pulse count in the negative
motion channel.

This mode of quick scanning is currently only possible for motors that
have associated encoder outputs.  At MR-CAT, only two motors have
this capability, namely the monochromator Bragg angle and the 8-circle
diffractometer's primary two theta angle.  However, SER-CAT plans to
make this possible for almost all motors on the beamline.
For SER-CAT, this is practical because they are using the Turbo PMAC
line of motor controllers from Delta Tau.
The Turbo PMAC is an up to 32 axis motor controller capable of driving
both stepper and servo motors, although SER-CAT is using only servo motors.
Turbo PMAC controllers are able to slave one of its motor axes to 
any of the other 31 motors controlled by it.  The slave axis can then
be commanded to generate a step and direction output which can be
converted by external electronics into the same kind of positive
and negative pulse trains as are used by MR-CAT.
Since the association of master and slave axes can be changed at run time,
this makes it possible to quick scan any of the other 31 motors controlled
by this PMAC.  While SER-CAT does not yet routinely operate in this mode,
a prototype of this configuration has already successfully been tested.

At present, most of the support for multichannel scalers in MX is for
the Struck SIS3801 MCS.  There is a small amount of support for alternate
MCS systems, but the main focus has been on the SIS3801.
The SIS3801 is a VME-based MCS that can have up to 32 input channels.
The SIS3801 acts as an MCS by transferring its measurements into a FIFO
that can hold up to 128K scaler values upon receipt of a channel advance pulse.
If all 32 channels are in use, this allows up to 4K measurements per channel.
However, the SIS3801 can be programmed to ignore some or even most of its
input channels, which then leaves more space for the remaining channels.
For example, if only 4 channels are used, then up to 32K measurements
per channel can be taken.

Currently, MX has two drivers for the SIS3801.  The first, called
\textit{epics\_mcs},
is for SIS3801s controlled by the EPICS MCA record
written by Mark Rivers of the University of Chicago
\cite{EpicsMCARecord}.
The second, called \textit{sis3801}, talks directly to the module
via MX VME I/O.  This configuration is intended for use with VME crates
that are controlled via PCI-to-VME bus interfaces or by MX servers
running directly on a VME crate controller.  The existence of two drivers
allows MX to support both VME crates controlled by EPICS and VME crates
that are not controlled by EPICS.
At a higher level, the two MX drivers
behave similarly.  The primary difference is that the \textit{epics\_mcs}
driver is currently restricted to a maximum of 4000 measurements per channel
due to a restriction in the maximum length of EPICS Channel Access messages.

The SIS3801 can be triggered to move to the next measurement by either
an internal or an external channel advance.  Use of the built-in 10 MHz
clock with the internal channel advance allows the SIS3801 to run
in a stand alone fashion.  However, sometimes it is desirable to synchronize
measurements by an SIS3801 with external devices.  For this reason,
MX also supports the use of the external channel advance signal.
Consequently, MR-CAT and SER-CAT have both purchased Struck SIS3807
pulse generator modules to serve as master clocks for the beamline.
Theoretically, one could synchronize the operation of multiple SIS3801s
in this fashion.

A more interesting use of the external channel advance support is for
synchronization with the MultiSCA firmware for the XIA DXP-2X MCA.
With this special firmware, the DXP-2X acts similarly to a multichannel scaler
in that each channel buffers a new set of 16 ROI integral measurements
upon receipt of an external channel advance signal.  With a common
external clock such as the SIS3807, it will be possible to implement
quick scans that also record MCA measurements for each point of
the quick scan.  Although some preliminary work has been done, this mode
of operation has not yet been fully implemented by MX yet.
However, we hope to be able to provide this mode of operation at MR-CAT
by the end of 2003.

\section{Performance}

The most important current task in MX development is improvement in
the performance of pieces of MX that are perceived to be slow.
When examined in detail, most of these issues revolve around
performance problems with the default MX network protocol.
We are currently addressing this issue in more than one way.

First of all, one possible solution is to use a different network protocol.
In an APS environment, the most obvious alternative is the
EPICS Channel Access protocol.  The author originally was concerned
about its dependence on VME crates and an expensive real time operating
system.  However, recent changes in EPICS licensing and the porting
of the server side of EPICS IocCore to Linux, Solaris, Win32, and RTEMS
have gone a long way toward mitigating these concerns.  MX can already
be used as an EPICS client via MX drivers such as \textit{epics\_motor},
\textit{epics\_scaler}, and \textit{epics\_mcs}.  We have also begun
work on making the MX drivers and toolkit available from the
EPICS IocCore server which is described further in Section VII.

We are also pursuing improvements to the existing MX network protocol.
It has been suggested that much of the overhead may be due to the
formatting and parsing of data sent over the network and due to the
lack of caching of the identity of remote records.  The author will
be focusing his efforts on this and similar ideas for the rest of 2002
and on into 2003.  Some of the testing will revolve around improving
the performance specifically of network XIA MCA readout,
since this makes particularly heavy use of the network.

Perceived responsiveness is also important for user satisfaction
\cite{GUIBloopers}.
Early versions of Imcagui, the main user GUI for crystallography,
had a beamline status update routine that could take around a second
to run.  This meant that if the user typed or clicked something on
the screen, it might take up to a second for the GUI to acknowledge this.
When the status update routine was modified to force frequent updates
of the screen, the delay in responding to user input dropped down to
a small fraction of a second.  The users were much happier with this
version of the GUI even though it was really doing work at about
the same rate as before.

\section{New and Upgraded Beamlines}

A significant amount of the recent work in MX has been in porting
the software to new beamlines.  Table \ref{tab:Beamlines}
lists the beamlines where MX is currently in routine use and
highlights of the equipment used there.  One thing that should be
apparent from Table \ref{tab:Beamlines} is that MX is able to support
a heterogeneous mix of equipment.

Much of the new work has been in support of the SER-CAT sector (Sector 22)
at the Advanced Photon Source.  SER-CAT uses almost exclusively 
Delta Tau Turbo PMACs for motion control which are controlled via RS-422
by several Linux-based MX servers.
Struck SIS3801 multichannel scalers are used for counter/timer support.
They are controlled from a National Instruments PCI-to-VME bus interface
via MX drivers.

\begin{figure}
\includegraphics{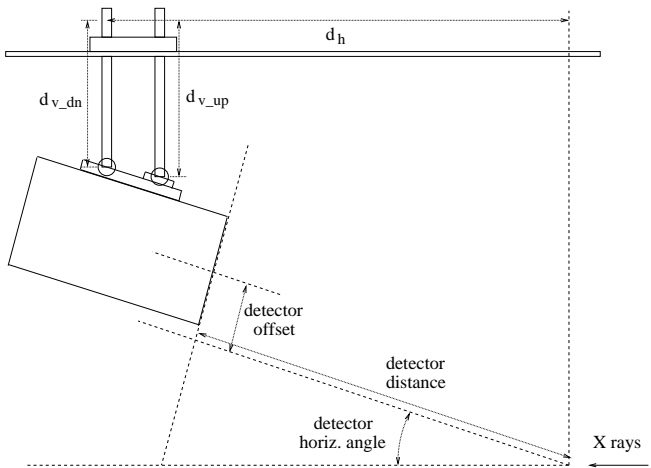}
\caption{\label{fig:AFrame} The A-frame detector support for SER-CAT.}
\end{figure}
SER-CAT beamline specific work has included the development of pseudomotors
to handle A-frame detector supports of the type originally developed for
SBC-CAT as shown in Figure \ref{fig:AFrame}.
These pseudomotors transform the vertical and horizontal motions of
the detector into detector two-theta angle, detector distance,
and detector offset.  Other work has included the development of software
to control the PMAC-based beamline goniostat from the MarCCD and
Bruker Proteus CCD systems via MX.

At IMCA-CAT,
a recent development has been support for the Rigaku/MSC sample changing
robot derived from the original Abbott design.  This task and support
of the Bruker Proteus system at SER-CAT has motivated the development
of a simplified server controlled via ASCII commands.  The idea here is
to make it as easy as possible for vendors to interface their systems
to the beamline.

A recent addition has been the new DND-CAT protein crystallography beamline
at APS Sector 5-BM.  This beamline implements low level hardware
control using their own protocol called SCIPE
\cite{SCIPEreference}.  MX applications are supported on this beamline
via a set of MX devices drivers such as \textit{scipe\_motor},
\textit{scipe\_scaler}, and so forth that are analogous to the ones used
to communicate with EPICS controlled devices.  Another new user of MX
is the Gulf Coast Protein Crystallography Consortium at CAMD.  They
are operating their beamline with Compumotor 6K motor controllers,
Ortec 974 counter/timers, and XIA Saturn MCAs.

One task that will become important in the near future will be more
complete automation of crystallography data collection.  The ideal
will be to make it so that crystals can be shipped to the beamline
in standardized containers.  These will then be automatically mounted
and aligned, followed by CCD data acquisition and processing.
The plans for how this will be implemented are not complete yet, but
SER-CAT plans to begin working on this very soon.  In fact, some early
steps have already been taken such as the purchase of x-y sample
positioning stages from Oceaneering Space Systems that are to be used
as part of automatic crystal alignment.

The first phase for SER-CAT will involve development 
of a standard interface for interaction with the remote experimenter,
including a method for transmitting images of diffraction patterns for remote 
evaluation.  At first, the staff will do the work, and then transmit the 
results.  Later, more real-time interaction will be developed.  This 
will probably need to be in the form of a CGI to allow the interaction 
and authentication.
Control will come later, and is actually easier. The other push will be
the development of robotics for automation of alignment and sample 
changing, which will require motion control programming.

Another aspect of the automation is the development of automatic
beamline alignment and optimization scripts, further automation of the 
fluorescence measurement, automation of data collection strategy for 
multiaxis diffractometry, and automated data reduction.

\section{MX as Device Support for EPICS}

A recent major development in the EPICS world has been the porting of the
EPICS server software, called IocCore, to Linux, Solaris, Win32, and RTEMS.
This is in beta releases of what is to become EPICS release 3.14.  Previous
versions of IocCore only ran on VxWorks.  However, the device drivers
for previous versions of IocCore were generally also VxWorks specific and
most have not been ported.  This means that the workstation version of
IocCore has relatively few device drivers at this point.
One of the design goals for MX has been to make it easily embeddable
in other packages.  Thus, an interesting idea would be to embed MX inside
of EPICS IocCore.  The EPICS motor record now has a partially finished
port to EPICS 3.14, so it is an ideal test case for this.  Correspondingly,
I am now working with Ron Sluiter to develop such an interface.

Since MX presents the same abstract high level API for all of its motor drivers,
my plan is to essentially treat MX as an EPICS driver support package.
Then, all that remains to be done is to create a device support module for
the EPICS motor record that translates requests from EPICS into a form that
MX can understand and then translate MX's responses back into a form
that EPICS can understand.  I believe that there is a good chance that this
can be completed quickly and hope to have a working version by the end of 2002.

The benefits of implementing such an interface are:
\begin{itemize}
\item[1.]  The EPICS MEDM program is a good static interface builder.
		This would allow MEDM to be used with MX-controlled motors.
\item[2.]  This would allow Spec to operate MX-controlled motors.
\item[3.]  MX has a fairly large motor driver library which currently
		contains 34 motor drivers and 17 pseudomotor drivers.
		These would all then be available more or less immediately
		to the rest of EPICS.
\end{itemize}
The initial development work for this will be done on Linux.
Once it works well,
it will then be ported to Solaris, Win32, and maybe even RTEMS.

\section{Conclusions}

In summary, much progress has been made in the last couple of years.
MX is now a full featured beamline control system for materials science,
protein crystallography, and other experiments.
It is expected that many new features will be added as needed to support better
beamline automation and new experimental techniques.

The author would like to thank MR-CAT, IMCA-CAT, and SER-CAT for the support
they have provided during the development of this software.
MR-CAT related development was supported in part by funding from the
U.\ S.\ Department of Energy under grant number DEFG0200ER45811.
IMCA-CAT related development was supported by the companies of the
Industrial Macromolecular Crystallography Association through a contract with
Illinois Institute of Technology (IIT), executed through the IIT's
Center for Synchrotron Radiation Research and Instrumentation.
SER-CAT related development was supported by the institutes of the
South East Regional Collaborative Access Team through a contract with
the University of Georgia.
Use of the Advanced Photon Source was supported by the 
U.\ S.\ Department of Energy, Basic Energy Sciences, Office of Science,
under Contract No.\ W-31-109-Eng-38.

\end{document}